\documentclass{article}
\usepackage{spconf,amsmath,graphicx, amsfonts}
\usepackage{booktabs}
\usepackage{tabularx}

\def\loss{{\mathcal L}}
\def\R{{\mathbb R}}

\title{BERT FOR JOINT MULTICHANNEL SPEECH DEREVERBERATION WITH SPATIAL-AWARE TASKS}
\name{Yang Jiao}
\address{University of Maryland, College Park, Maryland, USA\\
yjiao1@umiacs.umd.edu}
\begin{document}
%
\maketitle
\begin{abstract}

We propose a method for joint multichannel speech dereverberation with two spatial-aware tasks: direction-of-arrival (DOA) estimation and speech separation. The proposed method addresses involved tasks as a sequence to sequence mapping problem, which is general enough for a variety of front-end speech enhancement tasks. The proposed method is inspired by the excellent sequence modeling capability of bi-directional encoder representation from transformers (BERT). Instead of utilizing explicit representations from pretraining in a self-supervised manner, we utilizes transformer encoded hidden representations in a supervised manner. Both multichannel spectral magnitude and spectral phase information of varying length utterances are encoded. Experimental result demonstrates the effectiveness of the proposed method. 

\end{abstract}
\section{Introduction}
\label{sec:intro}

Reverberation is inevitable in an enclosed room space where multipath sound signals reflect at walls, floors and obstacles superpose at the receiver \cite{naylor2010speech}. This phenomenon degrades performance of indoor hearing aid and automatic speech recognition (ASR) systems. While reverberation falls into the category of convolutive noises, interfering speaker falls into the category of additive noises \cite{makino2007blind}. Reverberated source separation is more challenging than anechoic ones \cite{togami2020joint}. Besides speech dereverberation and separation, DOA estimation is a highly-demanded task under a multichannel setting. Again, the performance of conventional DOA estimation method degrades with the presence of reverberation. False spurs resulting from strong reverberation can always confuse conventional methods such as SRP-PHAT \cite{dibiase2000high} and MUSIC \cite{schmidt1986multiple}. Those tasks are widely studied separately; a joint approach is desirable and less studied. 

Recently, speech separation has been addressed with neural network in time domain with TasNet \cite{luo2018tasnet, luo2019conv}. Tasnet is suitable for realtime applications which favor low computation overhead and short processing latency. Taking another track, a variety of speech enhancement tasks have been extensively addressed with neural masks in the time frequency domain\cite{heymann2016neural, williamson2015complex, tan2018convolutional}. Compared with time domain approaches, those methods decouple spectral phases from spectral magnitudes. As a result, the reconstruction of waveforms has been a nontrivial challenge. Our work takes the time frequency domain track and models a complete utterance of varying length. We utilize compact temporal-spectral features, allow magnitudes and phases to be encoded; and address reconstruction with a pretrained neural vocoder \cite{yamamoto2020parallel}. Customized transformer and attention mechanism has also been studied in the speech enhancement context \cite{kim2020t, hao2019attention}; however a method general enough for a variety of tasks is less studied. We also keep the stacked transformer layers of moderate size, comparable to a single layer BLSTM. 

BERT, first proposed in natural language processing (NLP) \cite{devlin2018bert} for self-supervised representation learning, has achieved huge success in improving a wide range of downstream tasks. BERT has been introduced to self-supervised speech representation learning in \cite{liu2020mockingjay}. Compared with bi-directional long short-term memory (BLSTM), transformer network allows flexible context range and enables integrating utterance-level information. Transformer encoder is parallelizable, enabling deeper network to be trained. We apply several techniques to adapt BERT from NLP to speech, including downsampling \cite{pham2019very}, pre-sequence mapping \cite{pham2019very}, prenorm \cite{wang2019learning} and T-GSA weighting\cite{kim2020t}. 

In this paper, we consider two spatial-aware tasks jointly with dereverberation. In the first task, there is no interfering speaker. The model is trained to predict the DOA of the target speaker jointly with the prediction of anechoic spectral magnitude. This setup is inspired by \cite{mack2020signal, zhang2019robust, varzandeh2020exploiting}, where it is shown that joint consideration of voice-activity-detection (VAD) improves the performance of DOA estimation. In the second task, the task speaker's DOA is fixed; a second speaker is present at different DOAs. The task is to predict anechoic spectral magnitude for both target and the interference speaker. The permutation confusion is implicitly resolved by the neural network using the different spatial information of the speaker, thus no permutation invariant training (PIT) \cite{yu2017permutation} is applied.

\section{PROBLEM FORMULATION}
\label{sec:format}

We consider a multiple channel speech dereverberation problem. In the first task, the DOA of target signal $x(t)$ is denoted as $\alpha$. 
The signal received at microphone $m$ in time frequency domain $Y_m$ is given by

\begin{equation}
  Y_m(n, k) = H_m(n, k) \cdot X_m(n, k)
\end{equation}

where $ m = 1, ..., C, n = 1, ..., T, k = 1, ..., K$. $H_m$ and $X_m$ represent room transfer function from target speaker to microphone $m$ and target signal respectively. 
The goal is to estimate $x(t)$ and $\alpha$ jointly. 

In the second task, the signal received at microphone $m$ is given by 
\begin{align}
  Y_m(n, k) &= H_m(n, k) \cdot X_m(n, k)\\
   &+ T_m(n, k) \cdot Z_m(n, k)
\end{align}
where $T_m$ and $Z_m$ represent room transfer function from the interference speaker to microphone $m$ and interference signal respectively. The interference signal in the time domain is denoted as $z(t)$. 

The goal is to estimate $x(t)$ and $z(t)$ jointly. 

\section{Proposed Method}
\label{sec:pagestyle}

\begin{figure}[tb]
  \centering
  \includegraphics[width=\columnwidth]{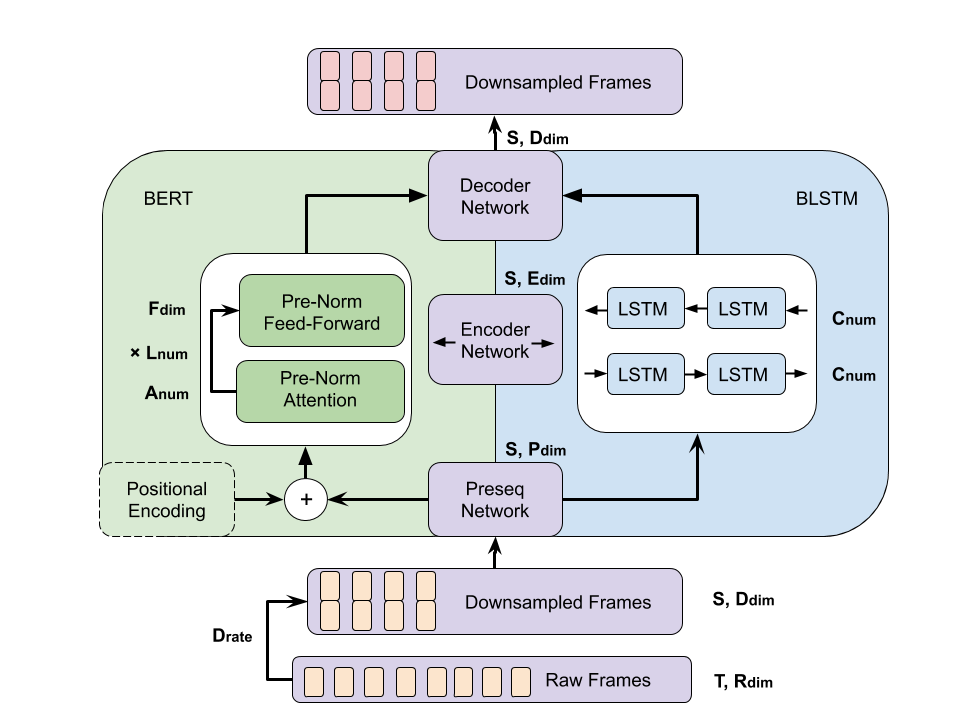}
  \caption{Overview of proposed framework. On the left: BERT encoder in green shade. On the right: BLSTM encoder in blue shade.}
  \label{fig:sys}
  \end{figure}

The system consists of following modules: downsample, preseq network, sequential network, decoder, generator. 
In both tasks, the input tensor $Y \in \R ^{C\times T \times F}, F=160$ is given by 
\begin{align}
  Y &= concat(mag, phase) \\
  mag &= logmel(Y_m(n,k = 1, ..., K)) \\
  phase &= angle(Y_m(n, k = 1, ..., 80))
\end{align}
where $ mag \in \R ^{C\times T \times 80}, phase \in \R ^{C\times T \times 80}$. $concat()$ is along the last dimension, $angle()$ is taking the phase of complex number, $logmel()$ is taking logarithm Melscale spectrogram of $80$ bins. 

The downsample operation stacks neighboring $dr=3$ frames to form a super-frame, and shortens the original sequence by $dr$. Then, $F$ bins of $C$ channels are concatenated. Then, we obtain a tensor $Y'$ of size $(T/dr, F\cdot dr\cdot C)$ as input to the preseq network and the rest of the network. 

In the first task, the generator produces two tensors $doa \in \R ^{T/dr \times cls}, cls=72$ and $mag' \in \R ^{T/dr \times F\cdot dr \cdot C}$. The prediction of DOA is obtained by voting over all frames. The loss to train the network is given by 
\begin{align}
  \loss &= l_2(mag', mag\_true) \\
  & + cross\_entroy(doa, doa\_true) \\
  mag\_true &= downsample(logmel(X_m))
\end{align}
where both $l_2$ loss and cross entropy loss is computed frame wise and then summarized for optimization. 

In the second task, the generator produces two tensors $mag0' \in \R ^{T/dr \times F\cdot dr \cdot C}$ and $mag1' \in \R ^{T/dr \times F\cdot dr \cdot C}$. The loss to train the network is given by 
\begin{align}
  \loss &= l_2(mag0', mag0\_true) \\
  &+ l_2(mag1', mag1\_true) \\
  mag0\_true &= downsample(logmel(X_m)) \\
  mag1\_true &= downsample(logmel(Z_m))
\end{align}
where $mag0$, $mag1$ stands for target speaker and interference speaker respectively. 

\subsection{NETWORK SPECIFICATION}

The preseq network is a fully-connected layer with nonlinearity that operates on the last dimension of the input sequence $Y'$. The input is projected to a space that is suitable for direct addition with positional encoding \cite{vaswani2017attention}

The sequential network is a stack of three transformer layers. We apply a sinusoidal positional encoding since transformer layer is invariant to input positions. We apply a prenorm variant of the transformer layer \cite{wang2019learning}, which doesn't block the flow of gradient at the layer normalization module. 

The attention layer is a variant that only take the absolute value of scores and apply exponentially decay weight to farther away frames. We take initial value of $\sigma = 10$ empirically \cite{kim2020t}.

Fig \ref{fig:sys} illustrates a system diagram. The selection of hyperparameter is summarized as \small $D_{rate}=3, F_{dim}=2048, L_{num}=3, A_{num}=8, D_{dim}=768, C_{num}=1024$. \normalsize

\begin{table*}[tb]
  \centering
\begin{tabular*}{0.85\textwidth}{@{}cccc||cccc@{}}
\toprule
Method   &T60 & Top5 1-Accuracy & Top5 MAE & Method &T60 & Top5 1-Accuracy & top5 MAE\\
\midrule
SRP-PHAT &0.3 & 52.10\%         & 10.807   & MUSIC  &0.3 & 61.34\%         & 
17.361 \\
         &0.6 & 51.58\%         & 14.863   &        &0.6 & 63.16\%         & 
21.705 \\
         &0.9 & 54.31\%         & 15.112   &        &0.9 & 65.52\%         & 
16.310\\ 
\bottomrule
\end{tabular*}
\label{tab: doa}
\caption{Top-5 misclassification rate (1-Accuracy) and mean absolute error (MAE) of SRP-PHAT and MUSIC. }
\end{table*}

\begin{table*}[tb]
  \centering 

  \begin{tabular*}{0.95\textwidth}{@{}ccccc||ccccc@{}}
  \toprule
  Speaker &T60 & Proposed       & BLSTM      & Unprocessed & Speaker &T60 & Proposed       & BLSTM      & Unprocessed   \\ \midrule
  spk1 &0.3 & 1.82/0         & 1.76/0     & 1.24/-7.38  &spk2&0.3 & 1.66/0         & 1.60/0     & 1.12/-5.88      \\
  &0.6 & 1.66/0         & 1.60/0     & 1.14/-6.57    & &0.6 & 1.53/0         & 1.48/0     & 1.10/-5.39      \\
  &0.9 & 1.53/0         & 1.49/0     & 1.12/-6.04   & &0.9 & 1.45/0         & 1.40/0     & 1.10/-6.64      \\ \bottomrule  
  
  \end{tabular*}
  \label{tab:task2-pesq}
  \caption{PESQ and SI-SDR (dB) of transformer based method, BLSTM based method, unprocessed. Speaker 1 has a fixed DOA at 0 degree. Speaker 2 has a varying DOA from 30 to 330 degree. }
  \end{table*}

  \begin{table*}[tb]
    \centering
    \begin{tabular*}{\textwidth}{@{}*{13}{c}@{}}
    \toprule
    Type & T30 & Unprocessed & No T-GSA & 0.2 & 1 & 10 & Type & T30 & No T-GSA & 0.2 & 1 & 10 \\
    \midrule
    Default & 0.3 & 1.55        & 1.95     & 1.78 & \textbf{2.02} & 1.99 & Dense & 0.3 & 1.83 &	1.78 &	1.97	& \textbf{2.01} \\
    15.22 M& 0.6 & 1.29        & \textbf{1.77}     & 1.40 & 1.72 & 1.72  & 13.05M& 0.6 & 1.44	 &1.40&	1.70&	\textbf{1.71} \\
    & 0.9 & 1.21        & 1.44     & 1.28 & 1.54 & \textbf{1.56}  && 0.9 & 1.30	&1.28	&1.55	& \textbf{1.56} \\
     \bottomrule
    \end{tabular*}
    \label{tab: sigma}
    \caption{PESQ of default and densely connected transformer layers of different initial sigma values. }
    \end{table*}

\section{EXPERIMENTS}
\label{sec:typestyle}
\subsection{DATASET}
To evaluate the proposed method, we create a synthetic dataset from the 3rd CHiME Challenge Dataset and a RIR simulator based on image-method \cite{lehmann2008prediction}.
The 3rd CHiME Challenge Dataset \cite{barker2015third} contains 7138 utterances in 'tr05\_org', 410 utterances in 'dt05\_bth' and 330 utterances in 'et05\_bth', which are original clean WSJ0 data, booth recorded data and booth recorded data respectively. For both tasks, the reverberation room is of size [4, 4, 2.5] meters. We use a circular microphone array with 4 microphones, 1 meter radius, sitting at the center of the room. In the first task, the position of the target speaker is uniformly sampled over a circle of 1.5 meter radius, with a resolution of 5 degree. In the second task, the position of the target speaker is fixed at degree 0; the position of the interference speaker is uniformly sampled with a 30 degree resolution from 30 degree to 330 degree. We uniformly sample over 3 reverberation times [0.3, 0.6, 0.9] seconds. In the second task, the SNR is 0dB by setting the maximum amplitude of both speakers the same. Each utterances in the tr/dt/et set is convolved with a uniformly sampled RIRs from all reverberation times and DOAs. 

\subsection{PREPROCESS, METRIC AND OPTIMIZATION}
We apply STFT with a Hanning window of 1200 points, FFT of 2048 points and a hop size of 300 points. The $logmel()$ function is operated over 80 to 7000 Hz with 80 mel bins. The feature is normalized to zero mean and unit variance \cite{yamamoto2020parallel}. 
We use PESQ \cite{rix2001perceptual} and SI-SDR \cite{le2019sdr} to evaluate our method. The network is trained with Adam optimizer. The learning rate is warmed up over the first 1\% of total 75k steps to a maximum of 3e-4 and then linearly decayed to 0. 

\subsection{TASK 1}

Table \ref{tab: task1-pesq} compares different approaches at various reverberation times. Mutiple-channel WPE \cite{drude2018nara, nakatani2010speech} is able to achieve the highest PESQ across all approaches when the reverberation is not severe. For severe reverberation, transformer and BLSTM based approaches outperform multichannel WPE. For all conditions, transformer based approach outperforms BLSTM based approach. Table \ref{tab: doa} compares non-neural approaches SRP-PHAT \cite{dibiase2000high} and MUSIC \cite{schmidt1986multiple}  DOA estimation on the evaluation set \cite{scheibler2018pyroomacoustics}. Neural transformer based approach misclassifies 1 out of 330 evaluation utterances. We can observe a significant performance degradation on non-neural approaches. To count for false spikes due to strong reverberation path, Top-5 metrics are listed along, which are metrics that consider the best performance among the 5 most probable DOAs given by the algorithm. The superior misclassification rate of transformer based method indicates its effectiveness in extracting spatial information under adverse conditions. 
\begin{table}[]
  \begin{tabular}{@{}cccccc@{}}
  \toprule
  T60                  & Proposed & BLSTM  & WPE-1  & WPE-4  & Unproc     \\ \midrule
  0.3                  & 2.24/0   & 2.18/0 & 1.72/- & \textbf{2.82}/- & 1.59/-5.34 \\
  0.6                  & 2.03/0   & 1.95/0 & 1.32/- & \textbf{2.13}/- & 1.28/-5.35 \\
  0.9                  & \textbf{1.83}/0   & 1.76/0 & 1.23/- & 1.57/- & 1.21/-5.86 \\ \bottomrule
  \end{tabular}
  \label{tab: task1-pesq}
  \caption{PESQ and SI-SDR (dB) of transformer based method, BLSTM based method, WPE of single channel, WPE of 4 channels and unprocessed}
  \end{table}
\subsection{TASK 2}
\begin{figure}
\begin{tabular}{c}
\includegraphics[width=\columnwidth]{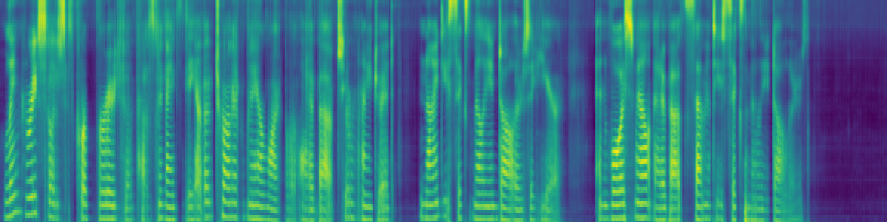} \\
(a) speaker1 processed \\[6pt]
\includegraphics[width=\columnwidth]{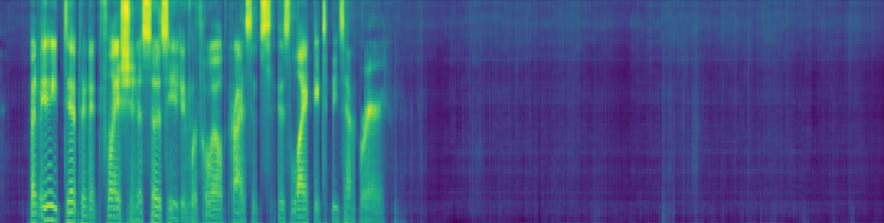} \\
(b) speaker2 processed \\[6pt]
\includegraphics[width=\columnwidth]{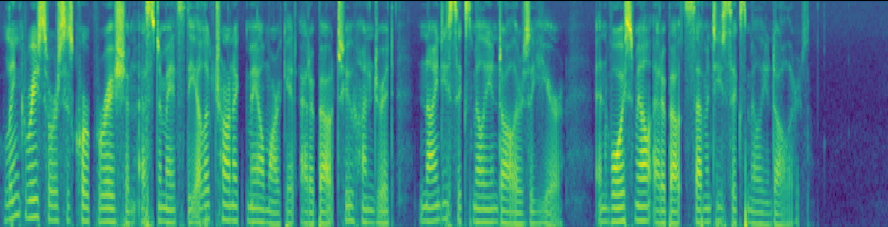} \\
(c) speaker1 clean \\[6pt]
\includegraphics[width=\columnwidth]{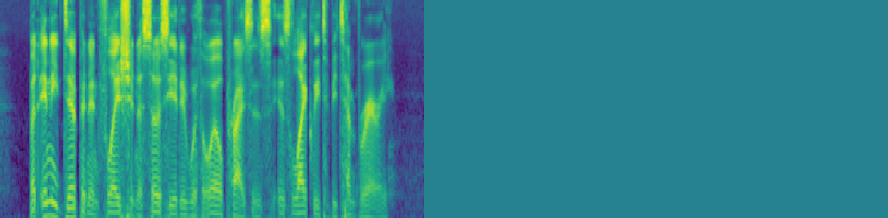} \\
(d) speaker2 clean \\[6pt]
\includegraphics[width=\columnwidth]{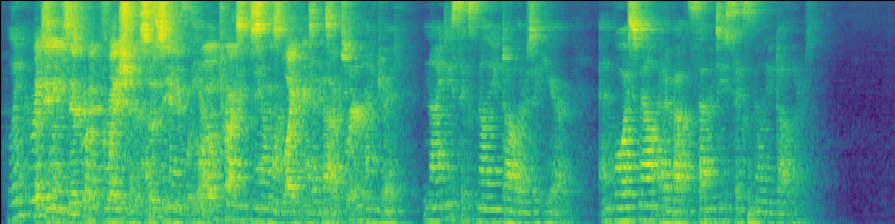} \\
(e) mixture unprocessed
\end{tabular}
\label{fig: task2}
\caption{Log-mel spectrogram before and after applying proposed method. }
\end{figure}

Table \ref{tab:task2-pesq} compares different approaches for the task of speech separation at different reverberation times. Transformer based approach outperforms BLSTM based method across all cases. For both metrics, we observe that the presence of an interference speaker poses the task a much more challenging one. 
To get a qualitative evaluation, we listen to the transformer processed samples in the evaluation set. The interfering speaker could not be heard and most degradation comes from dereverberation and reconstruction. Figure \ref{fig: task2} illustrates log-mel-spectrogram before and after processing. Although the mixture before processing provides only a smeared and superposed spectrogram, the network is able to extract anechoic spectrogram from it for each speaker. 

\subsection{SINGLE CHANNEL DEREVERBERATION}
To better understand the transformer based approach, we evaluate its performance on a single channel dereverberation task with different initial sigma values. Table \ref{tab: sigma} compares different sigma initial values. A larger value means the weight decays slower with distance and a larger context window as a result. A larger initial sigma value outperforms smaller values. The performance degrades when no weight is applied, due to the difficulty in optimization when all frames are considered. The performance also degrades when the initial signal value is smaller, when too few frames are considered. We also consider a densely connected transformer layer inspired by its CNN counterpart \cite{huang2017densely}. Comparable performance is achieved with less parameters; same pattern of initial sigma values is observed. 

\section{CONCLUSION}
\label{sec:conc}

In this paper, we investigated multichannel dereverberation jointly with direction-of-arrival (DOA) estimation or speech separation. We proposed a transformer network that maps contaminated frames to clean frames. Experiment results show that the transformer network is able to achieve significantly better PESQ and SI-SDR after processing and has the ability to perform task of DOA estimation or speech separation simultaneously.

\vfill\pagebreak

\bibliographystyle{IEEEbib}

{\footnotesize \bibliography{refs}}

\end{document}